**Denoising 2-D diffraction images by compressed sensing**

James Weng[1]*, Niklas B. Thompson[1], Christopher Folmar[2], James D. Martin[3], Christina Hoffmann[4]

1. Advanced Photon Source, Argonne National Laboratory

2. BWX Technologies Inc

3. Department of Chemistry, North Carolina State University, Raleigh, NC 27695-8204.

4. Oak Ridge National Laboratory.

**Abstract:**

**Intro:**

To obtain the best resolution for any measurement there is an ever-present challenge to achieve maximal differentiation between signal and noise over as fine of sampling dimensions as possible. In diffraction science these issues are particularly pervasive when analyzing small crystals, systems with diffuse scattering, or other systems in which the signal of interest is extremely weak and incident flux and instrument time is limited. We here demonstrate that the tool of compressed sensing, which has successfully been applied to photography,[i] facial recognition,[ii] and medical imaging,[iii,iv] can be effectively applied to diffraction images to dramatically improve the signal-to-noise ratio (SNR) in a data-driven fashion without the need for additional measurements or modification of existing hardware. We outline a technique that leverages compressive sensing to bootstrap a single diffraction measurement into an effectively arbitrary number of virtual measurements, thereby providing a means of super-resolution imaging.

In diffraction experiments, measurements are generally performed with an area detector which provides a 2D grid of measured values (intensity of photons impinging on the detector per pixel). For difficult to measure scattering effects, such as diffuse scattering, where long collection times are necessary for signal resolution, application of compressed sensing techniques can significantly speed up the collection of data with improved signal fidelity relative to conventional sampling methods.

A requirement for compressed sensing techniques to successfully recover some target signal is that the target signal be compressible, or representable in some basis in which it is mostly zeros. It can intuitively be understood that most natural images are compressible in the sense that they are non-random—that is, they contain structure.[v] A diffraction image necessarily contains some sort of structure, such as radial or translational symmetry, and thus contains *redundant information*. The simulated diffuse scattering pattern of $Hg(NH_3)_2Cl$ *(fig 1a)* for example, may be easily observed to be compressible by viewing its discrete cosine transform *(fig 1b)*. It is observed that the discrete cosine transform is a sparse representation of the original image (with the nonzero coefficients contained in ≈5% of the image area bounded by the red box). On the other hand, random noise remains non-compressible in any basis; from an information-theoretic perspective, random noise is *maximally information-rich* (SI #). This observation serves as the basis for the denoising algorithm presented below.

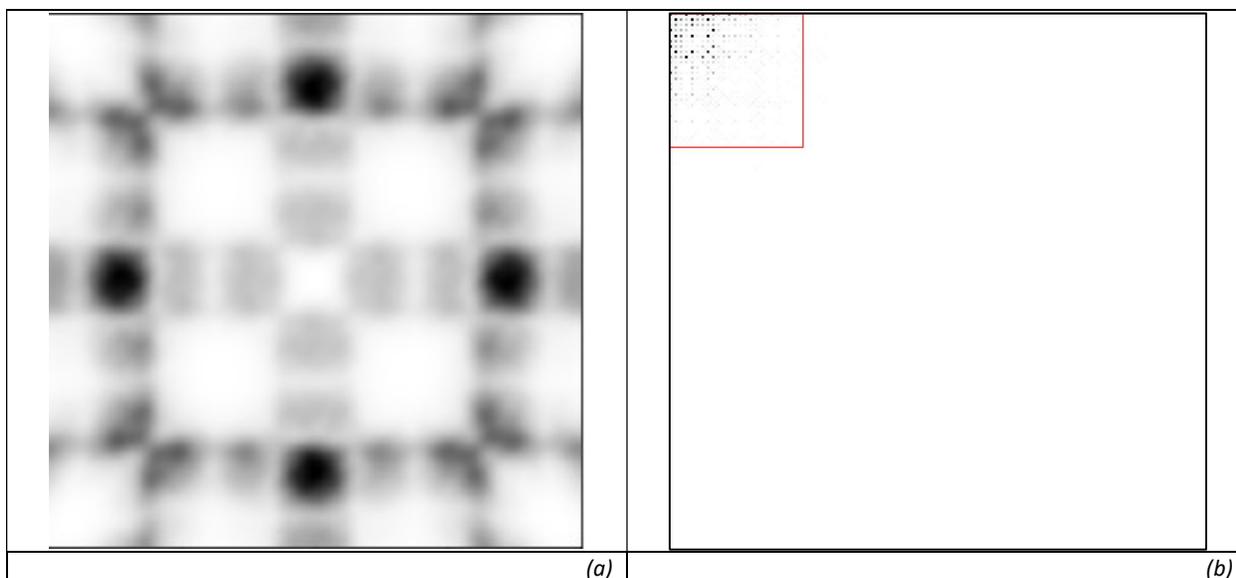

*Figure 1 : A simulated 2D diffuse scattering pattern of a model structure for Hg(NH$_3$)$_2$Cl[vi] (a) and its discrete cosine transform (b). The diffuse scattering pattern is sparse in the discrete cosine transform basis and is thus almost entirely empty space, with all nonzero coefficients being within ≈5% of the area bounded by the red box.*

**Introduction to compressed sensing:**

Ideally it would be possible to perfectly measure a signal of interest. However, in practice the signal can only be sampled with finite frequency (e.g., limited by the density of detector pixels), and with a limited experimental budget (e.g., limited time on an instrument). The successful measurement of a true underlying signal necessarily requires some prior knowledge about the signal being measured. To measure an arbitrary band-limited signal the conventional information-theoretic wisdom requires sampling at twice the highest frequency contained within the signal. This is referred to as the Nyquist-Shannon sampling theorem.[vii] For signals which are not band-limited, such as images, the sampling rate is generally dictated by the desired resolution. Thus for conventional measurements it is necessary to know the highest frequency in the signal being measured to determine the minimum sampling frequency. In practice, signals are often simply sampled at as high of a frequency as practical.

Diffraction patterns, which generally exhibit some level of symmetry, can intuitively be understood to have some level of redundant information. The prior knowledge that diffraction peaks have some specific profile[viii,ix] with dependence on scattering angle[x] has been used to great success for the fitting and extraction of structural information from diffraction patterns. This is the basis for Rietveld refinement,[viii,ix] originally developed to analyze low resolution neutron diffraction signals and illustrates that prior knowledge of the signal being measured allows the effective reconstruction of an under sampled, or low resolution, diffraction pattern.

Compressed sensing[xi], allows for signal reconstruction from measurements that are even more dramatically under sampled, relative to what is suggested by the Nyquist-Shannon sampling theorem. From an information theory perspective, a signal which is entirely random noise will have maximal information content. This signal, $x = [x_1, x_2, ..., x_N]$ where all $x_n$ are non-zero, cannot be perfectly represented by fewer than $N$ values and thus is incompressible. Though high in signal information content, such signals are in general uninteresting with respect to measurement information of scientific value. Conversely, a compressible signal is one in which there is some mathematical basis, $\psi$, in which the signal contains mostly zero values. That is, a compressible signal $x$ is one such that $x = \psi \cdot S$ where $S$ is sparse.

An example of a compressible signal is a 30 Hz sine wave, which in the Fourier basis is nearly all zero values. In a scientific context, a measured oscillating signal provides some scientific information about a physical phenomenon, while being low in information theoretic content as it is mostly zeros in the Fourier basis. In a real measurement, this oscillatory signal would contain random noise. As such the scientifically interesting and compressible signal, primarily not being composed of random noise, is substantially distinct from the incompressible surrounding noise *in structure*. Compressed sensing requires that the target signal be of low information theoretic content for a successful reconstruction. A diffraction pattern being a periodic function, it is a signal which, when represented in something like a Fourier basis, should be significantly compressible and thus should be well suited for compressed sensing signal reconstruction.

Ordinary measurements may be described by the equation $y = Ax$, where $y$ is the measured signal $y = [y_1, y_2, ..., y_N]$, A is the sensing (measurement) matrix, and $x$ is the true underlying signal of interest $x = [x_1, x_2, ..., x_N]$. This is represented by the following system of equations:

$$\begin{bmatrix} y_1 \\ y_2 \\ \vdots \\ y_N \end{bmatrix} = \begin{bmatrix} a_{00} & \cdots & & a_{0N} \\ & \ddots & & \\ \cdots & & \ddots & \cdots \\ a_{N0} & \cdots & & a_{NN} \end{bmatrix} \begin{bmatrix} x_1 \\ x_2 \\ \vdots \\ x_N \end{bmatrix} \quad \text{(Eqn. 1)}$$

In the case of a signal such as an image, where the collected signal is in the domain of interest, and $y = x$ (i.e., a perfect measurement), the sensing matrix is simply the $NxN$ identity matrix $I$. Compressed sensing can be described by an analogous equation $y = \Theta x$. However, y is now a compressed representation of the signal $y = [y_1, y_2, ..., y_M]$, $\Theta$ is an $NxM$ sensing matrix, $x = [x_1, x_2, ..., x_N]$ is the signal being measured. This is represented by the following system of equations:

$$\begin{bmatrix} y_1 \\ y_2 \\ \vdots \\ y_M \end{bmatrix} = \begin{bmatrix} \Theta_{11} & \Theta_{12} & \cdots & \Theta_{1N} \\ \Theta_{21} & \ddots & & \\ \vdots & & \ddots & \\ \Theta_{M1} & & & \Theta_{NM} \end{bmatrix} \begin{bmatrix} x_1 \\ x_2 \\ \vdots \\ x_N \end{bmatrix} \quad \text{(Eqn. 2)}$$

The matrix $\Theta$ is described by the equation $\Theta = \psi \cdot \Phi$ where $\Phi$ is an $MxN$ random measurement matrix and $\psi$ is an $NxN$ matrix which describes the basis in which $x$ is sparse. In expanded form, thus equation 2 becomes the following:

$$\begin{bmatrix} y_1 \\ y_2 \\ \vdots \\ y_M \end{bmatrix} = \begin{bmatrix} \Phi_{11} & \Phi_{12} & \cdots & \Phi_{1N} \\ \Phi_{21} & \ddots & & \\ \vdots & & \ddots & \\ \Phi_{M1} & & & \Phi_{MN} \end{bmatrix} \begin{bmatrix} \psi_{11} & \psi_{12} & \cdots & \psi_{1N} \\ \psi_{21} & \ddots & & \\ \vdots & & \ddots & \\ \psi_{N1} & & & \psi_{NN} \end{bmatrix} \begin{bmatrix} x_1 \\ x_2 \\ \vdots \\ x_N \end{bmatrix} \quad \text{(Eqn. 3)}$$

The compressed signal x is assumed to be $K$ sparse in some basis $\psi$. That is there are $K$ non zero elements, and $K << N$. A compressed sensing matrix is composed of M incoherent linear projections where $K < M << N$ and $\Theta^* \Theta \approx I$.

For example, consider some sensing system where the signal is collected in the Fourier domain. The sensing matrix A is the discrete Fourier transform matrix shown in Fig. 1a. Operation on this matrix by the measurement matrix $\Phi$, randomly (incoherently) deletes rows from this matrix (Fig. 2b) provides the sensing matrix $\Theta$ (Fig. 2d). Multiplying the sensing matrix $\Theta$ by its complex conjugate $\Theta^*$ (Fig. 2c) yields a matrix that is approximately an identity matrix with some small off diagonal terms (Fig 2e). This relationship between the sensing matrix and its complex conjugate allows the recovery of the desired signal from the compressed representation in equation 2 is a requirement for compressed sensing and is referred to as the restricted isometry property (RIP).[xii,xiii]

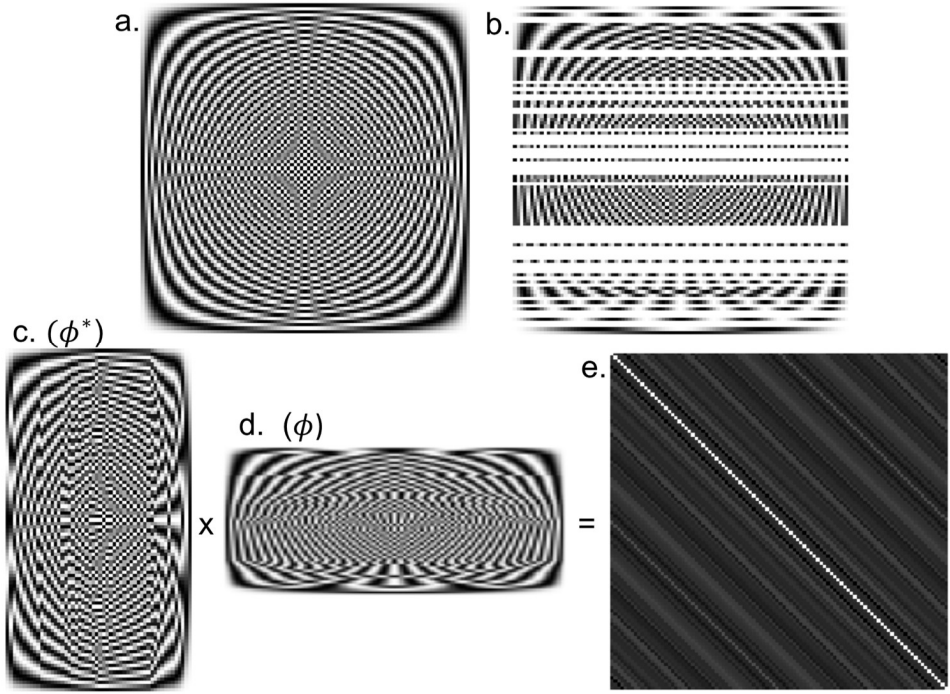

*Figure 2: Relationship between the sensing matrix and its compressed representation in a sparse basis. (a)The discrete Fourier transform matrix, which when multiplied by some signal x, results it the discrete Fourier transform of x. (b) Rows are randomly deleted from Fourier transform matrix to yield the compressed sensing matrix $\phi$ (d), and complex conjugate $\phi^*$ (c). (e) The product of a compressed sensing matrix ($\phi$) and its complex conjugate ($\phi^*$) return an approximate identity matrix with some small off diagonal terms. This property of the compressed sensing matrix and its complex conjugate allows for the recovery of a desired signal from its compressed representation.*

The measurement matrix $\Phi$ is arbitrary, and corresponds to randomly sampling the signal of interest and its sparsity will depend on the basis chosen. This random sampling is a necessary criteria for compressed sensing, similar to the minimum sampling rate defined by the Nyquist-Shannon sampling theorem for traditional measurements. For a real experimental signal, one does not know a priori the exact sparsity, $\rho = K/N$, of the signal, so the subsampling rate, $\delta = n/N$, is chosen to be some value expected to be greater than the actual sparsity, and the sensing matrix is constructed accordingly.

Specifically, in this work we consider diffraction patterns containing substantial diffuse scattering as the signal of interest. Empirically, these data all appear to be less than $\rho = 0.1$ sparse when represented as a discrete cosine or wavelet transform. As such, a subsampling rate of $\delta = 0.3$ is used (see the SI for empirical justification of this choice). Successful reconstruction in compressed sensing is governed by what is described as a compressed sensing phase transition diagram which identifies the probability of successfully reconstructing a signal that is randomly subsampled.[xiv] For a signal of sparsity $\rho = 0.1$, a subsampling rate $\delta = 0.3$ provides a probability of successful reconstruction very near 1. The sparser a signal is (less signal information), the fewer samples are necessary for a successful reconstruction. Conversely, more sampling is required to reconstruct a signal which is less sparse (higher signal information content). A completely random signal $N$, being sparse in no basis, requires $N$ samples to be correctly reconstructed.

The equation $y = \Theta x$, when $x$ is sparsely sampled, is thus an under-determined system of linear equations. The problem statement for compressed sensing is to find $x$ such that the sparsity of $x$ is enforced, stated as

$$min\ \|x\|_{l_0}\ subject\ to\ y = \Theta x \qquad (Eqn.\ 4)$$

where the goal is to minimize the total number of nonzero elements in $x$, the $L^0$ norm. This is classically difficult (NP-hard), but it is known that the $L^1$ norm is tractable heuristic for the $L^0$ norm, provided the RIP holds,[xv] thus the problem can be addressed as

$$min\ \|x\|_{l_1}\ subject\ to\ y = \Theta x \qquad (Eqn.\ 5)$$

Minimization of the $L^1$ norm is a commonly encountered problem in machine learning for which there are several existing computational methods.[xvi,xvii] In particular an algorithm known as Orthant-Wise Limited-memory Quasi-Newton (OWL-QN) that is based on the Limited-memory Broden-Fletcher-Goldfarb-Shanno algorithm (L-BFGS) provably converges to a globally optimal vector $x$.[xviii] In this work a python wrapper[xix] for a C implementation[xx] of an OWL-QN algorithm is used. For the minimization, the signal of interest is treated as an N-dimensional vector with no specific knowledge of its structure beyond the assumption that it is sparse; a $1000 \times 1000$ pixel diffraction image is handled as a 1,000,000-dimensional vector. Provided the subsampling rate $\delta$ is greater than the required subsampling rate for the signal sparsity $\rho$ as defined by the compressed sensing phase diagram, there is a high probability that the optimized signal x will be the accurate reconstruction of the actual sensing signal.

A stepwise tutorial for a toy compressed sensing reconstruction of a 1D signal of three overlapping sinusoids is presented in supporting information section 1. Generalized to any signal, compressed sensing reconstruction of a signal of size $N$ requires the following: (a) Incoherent (random) subsampling of a signal, (b) Transformation of that signal to a representation in a basis in which it is sparse. This provides $x$ in the system of equations described in equation 2, (c) Optimization of this system of equations 2 such that the $L^1$ norm of $x$ is minimized, thus approximating minimization of the $L^0$ norm of $x$. The minimization of $L^0$ of $x$ provides the sparsest representation of the signal for the information which was measured in the subsampling of the signal. (d) The optimized sparse representation is then transformed back to the domain of interest, in this example by taking the inverse discrete cosine transform.

**Application to diffraction:**

Returning to the simulated diffuse scattering pattern of Hg(NH3)2Cl (figure 1b), the area bounded by the red box is roughly 5% of the area, indicating that the original diffraction pattern is at least 95% redundant. This indicates that the signal has a sparsity of $\rho = 0.05$ at most. From the compressed sensing phase diagram any incoherent subsampling $\delta > 0.05$ should allow for reconstruction of the signal. Starting with the original image, 95% of the pixels may be discarded at random and from the subsampling a signal may be reconstructed *(fig 3)*. Multiple reconstructions of this test data are shown in the supplemental information, along with plots of the difference with the original input signal *(SI 1)*.

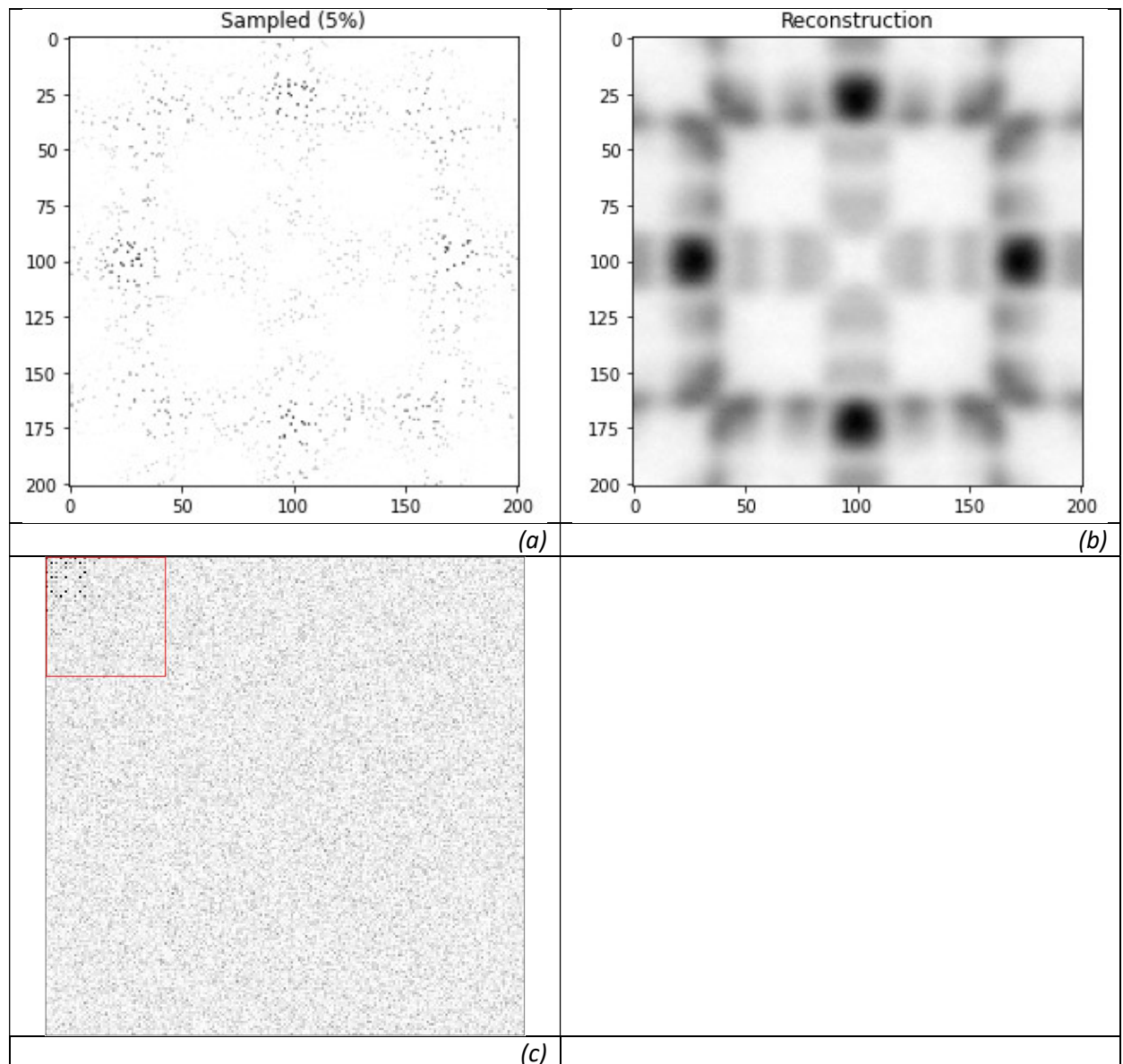

*Figure 3: A random subsampling of 5% of the pixels from simulated diffuse scattering of a model structure of Hg(NH$_3$)$_2$Cl (a) along with the resulting reconstruction from the subsampling (b). When viewing the discrete cosine transform of the diffuse scattering when random noise is added (c) it is evident that the added random noise is unstructured and distributed across the transform image, with the coefficients of interest still occupying the same location (bounded in red).*

When white noise is added to the input signal, because of its high information content, the representation of the overall signal is no longer sparse in the discrete cosine transform (fig. 3c) Additional examples of random noise are shown in the supplemental information *(SI 2)*.

It can be imagined that in an experimental diffraction experiment, some arbitrary function is sampled with an array of independent pixels, each with some measurement error/noise characteristics, illustrated below *(fig 4a)* as different colors. The measured value at each pixel may be described by the following equation:

$$I_m(x,y) = I_s(x,y) + \varepsilon(x,y) \qquad \text{(Eqn. 6)}$$

Where $I_m$ is the measured intensity, $I_s$ is the intensity of the signal of interest, $\varepsilon$ is the measured noise, and the pixel position is given by *(x, y)*. The detector can be considered to have some overall noise characteristic that influences the $\varepsilon$ terms contributing to the noise observed in the measurement; random variations in pixel responses created by manufacturing defects or radiation damage will alter the overall $\varepsilon$ of the detector. It is assumed that a flat field correction is applied to the detector, such that the $\varepsilon$ terms are not systematically biased in certain regions of the detector.

To reconstruct the signal of interest, which is sparse, most of the pixels are unneeded and may be discarded at random *(fig 4b)*. From here, an overall image may be reconstructed from the subsampling where the discarded pixels are reconstructed, by the problem stated in equation 5. Because $I_s$ is signal which is sparse in some basis, the original pixels $I_m$ contain redundant information about $I_s$ and it is possible to reconstruct $I_s$ from a subset of pixels. The noise $\varepsilon$ however, is high in information content and cannot be accurately reproduced from any subsampling. Attempted reconstructions of noise are shown in SI figure 2b. A reconstructed $I_m$ thus contains a faithful reconstruction of $I_s$ but an incorrect reconstruction of $\varepsilon$. Each reconstructed image can be thought of as having some different noise characteristic from the original measurement, with different noise content at the reconstructed pixels *(fig 4c)*.

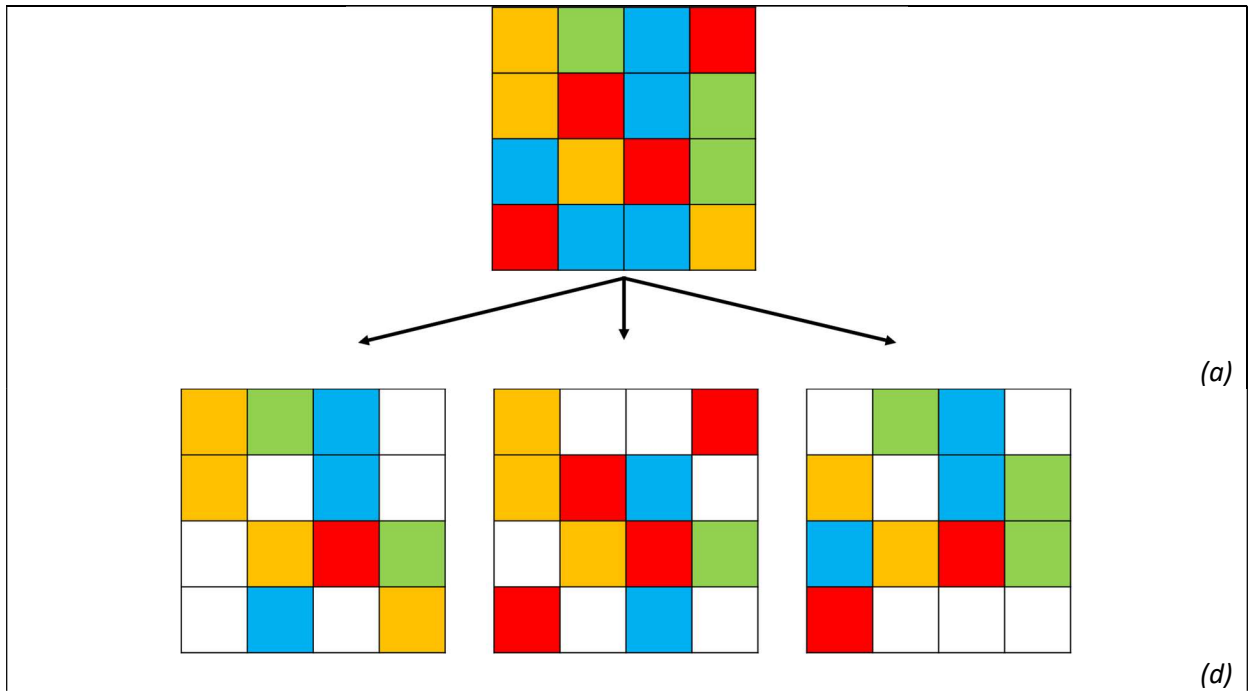

(a)

(d)

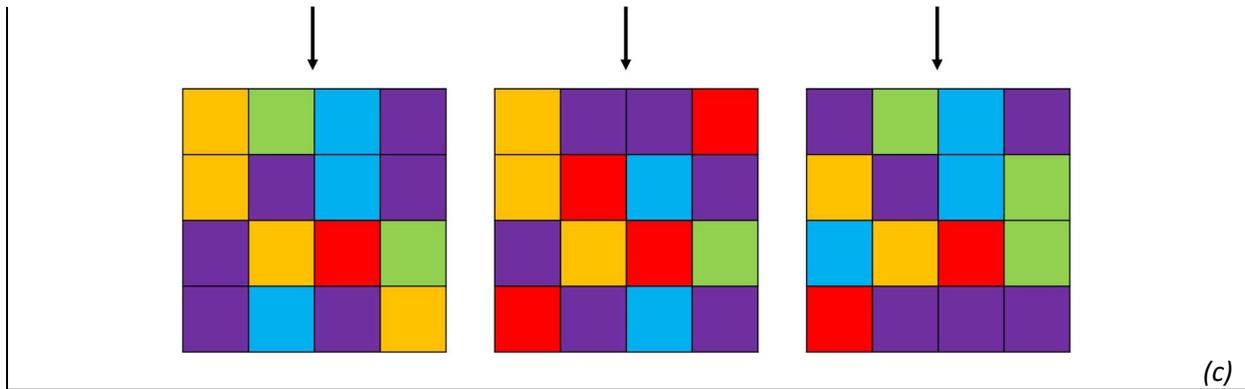

(c)

*Figure 4: An example of a detector which consists of an array of pixels, each with a different measurement error represented by a color (a). In a compressed sensing scheme, pixels may be randomly discarded (b) and reconstructed (c), effectively creating a new array of pixels with differing measurement error characteristics than the original array. Discarded pixels are shown in white, reconstructed pixels are shown in purple.*

This reconstructed signal can thus be thought of as being a new measurement on a different virtual detector with different overall noise characteristics than the original detector. This virtual detector has a lower overall amount of noise, because failure to reproduce $\varepsilon$ is the result of deleting non redundant portions of $\varepsilon$. By taking multiple different subsamples, a series of measurements on imaginary detectors may be computed, each with slightly different noise characteristics than the original measurement *(fig 4c)*. Because the signal of interest $I_s$ is highly redundant, while none of $\varepsilon$ is redundant, each reconstruction from a different subsets necessarily contain the signal of interest with added different improperly reconstructed noise.

This effectively allows one real detector measurement to be transformed into a very large number of virtual detector measurements. Averaging together many virtual detector measurements, the error term $\varepsilon$ in equation 6 would eventually coverge to some value much lower than the error in the original measurement, assuming that the original noise is unbiased and normally distributed.

The total number of different combinations of pixels can be calculated by the following equation:

$$C(n, r) = \frac{n!}{r!(n-r)!} \qquad \text{(Eqn. 7)}$$

Where *n* is the total number of pixels and *r* is the number of pixels used in a subsample. For a Perkin Elmer area detector commonly used in x-ray diffraction experiments with 2048x2048 pixels, the total number of combinations where 30% of the pixels are subsampled is ≈$10^{277576}$. However, the majority of these combinations likely do not meet the critera of pixels being "random" enough for a successful reconstruction, or are nearly identical to other subsamplings. It is likely that the average of virtual detector measurements converges to some value long before the total possible subsamplings is exhausted.

**Results:**

A noise free amorphous scattering signal with random ring placement and broadening was generated to have the ground truth signal, $I_s$, described in equation 6. Random noise was added to this signal in order to provide a "measured" signal with a known amount of noise *(fig 5a)*. The standard deviation of the noise content may be quickly estimated[xxi], providing an estimated σ = 50.1 which agrees with the expected σ = 50.

A 10% random subsampling where 90% of the pixels in the measured signal are discarded *(fig 6b)* and used for a compressed sensing reconstruction *(fig 5c)*. This subsampling fraction was arbitrarily selected. The noise in the reconstruction has an estimated σ = 30.7, the expected consequence of the inability to reconstruct noise. The reconstructed image also qualitatively appears to be less noisy.

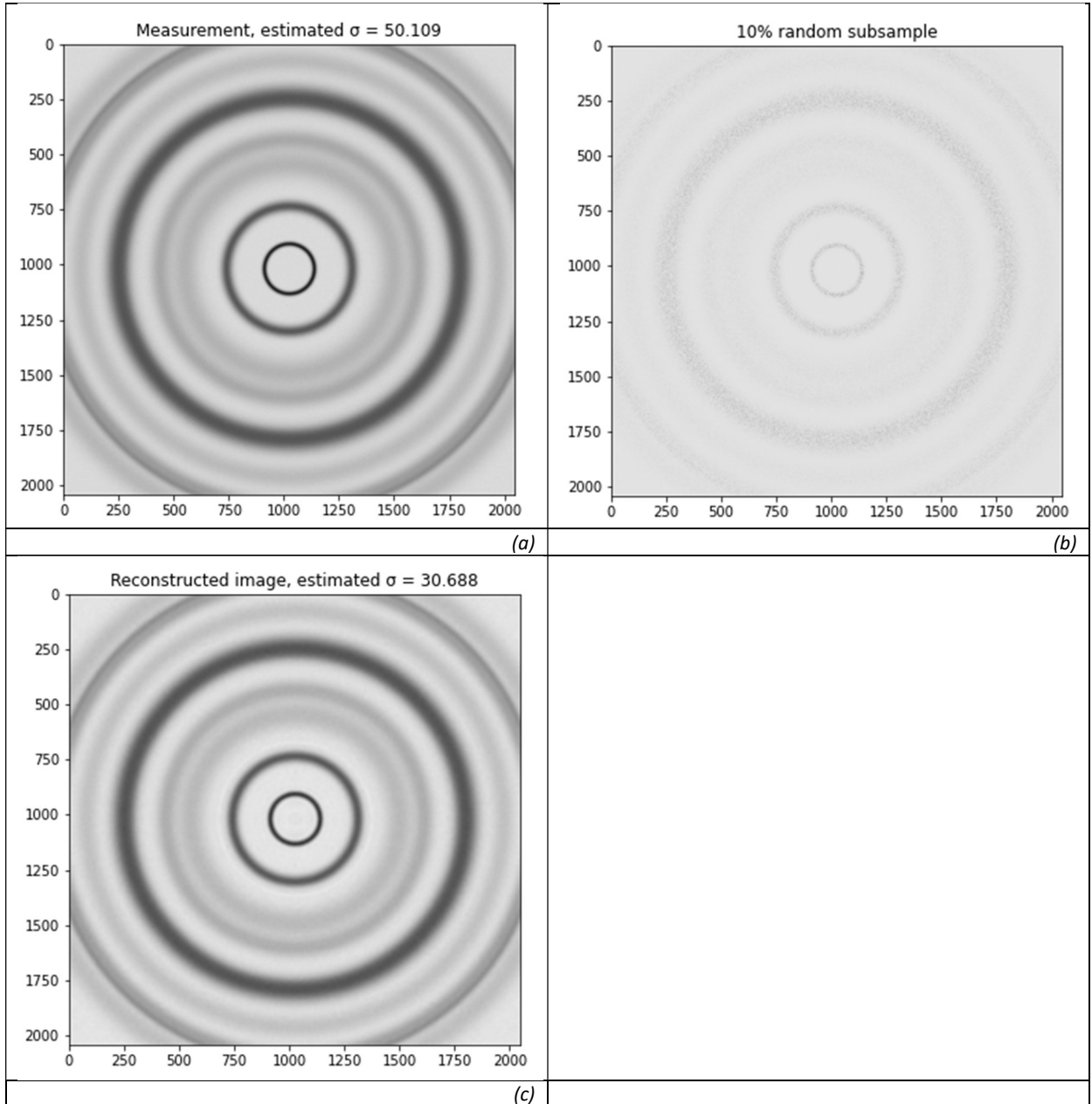

Figure 5: Generated amorphous scattering pattern with gaussian distributed noise added (a). A 10% subsampling of the pixels was taken (b) and a compressed sensing reconstruction was performed (c). The reconstructed image contains a smaller magnitude of added noise due to the inability of compressed sensing to correctly reconstruct noise.

Multiple 10% reconstructions are computed with 20 of them averaged together, producing an image with significantly reduced noise relative to the original measured data. Viewing a 1D trace through the averaged image *(fig 6)* shows a significant reduction in noise, as well as faithful reproduction of peak shapes for the average of the 20 compressed sensing reconstructions. More conventional noise reduction techniques, such as a median filter and a 3rd order Savitsky-Golay filter, show peak distortion and ringing effects which are not present in the compressed sensing reconstruction. Median and S-G filters were performed with a 25 point window.

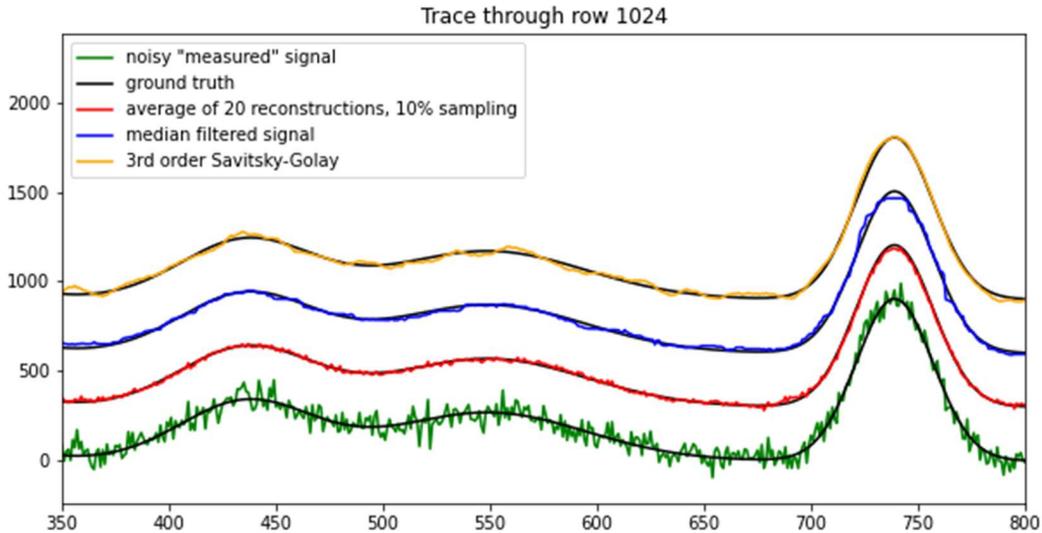

Figure 6: A 1D trace through the row 1024 of the simulated measured data, from pixel 350 to 800, showing several broad peaks. The noiseless ground truth is shown in black. The original data (green) is observed to be fairly noisy. The average of 20 reconstructions (red) is observed to closely match the known ground truth. Common denoising techniques such as a median filter (blue) and a 3rd order Savitsky-Golay filter (orange) are applied to the measured data as well, showing significant peak distortion and ringing effects. Window sizes for both the median and S-G filters are 25 pixels.

Addition of additional reconstructions to the average shows a decreasing trend in the estimated $\sigma$ of the noise *(fig 7a)*. This trend also follows when looking at the sum of absolute errors (SI Figure 3) between the averaged reconstructions and the known ground truth. An identical trend is seen when simply averaging frames of generated random noise *(fig 7b)*.

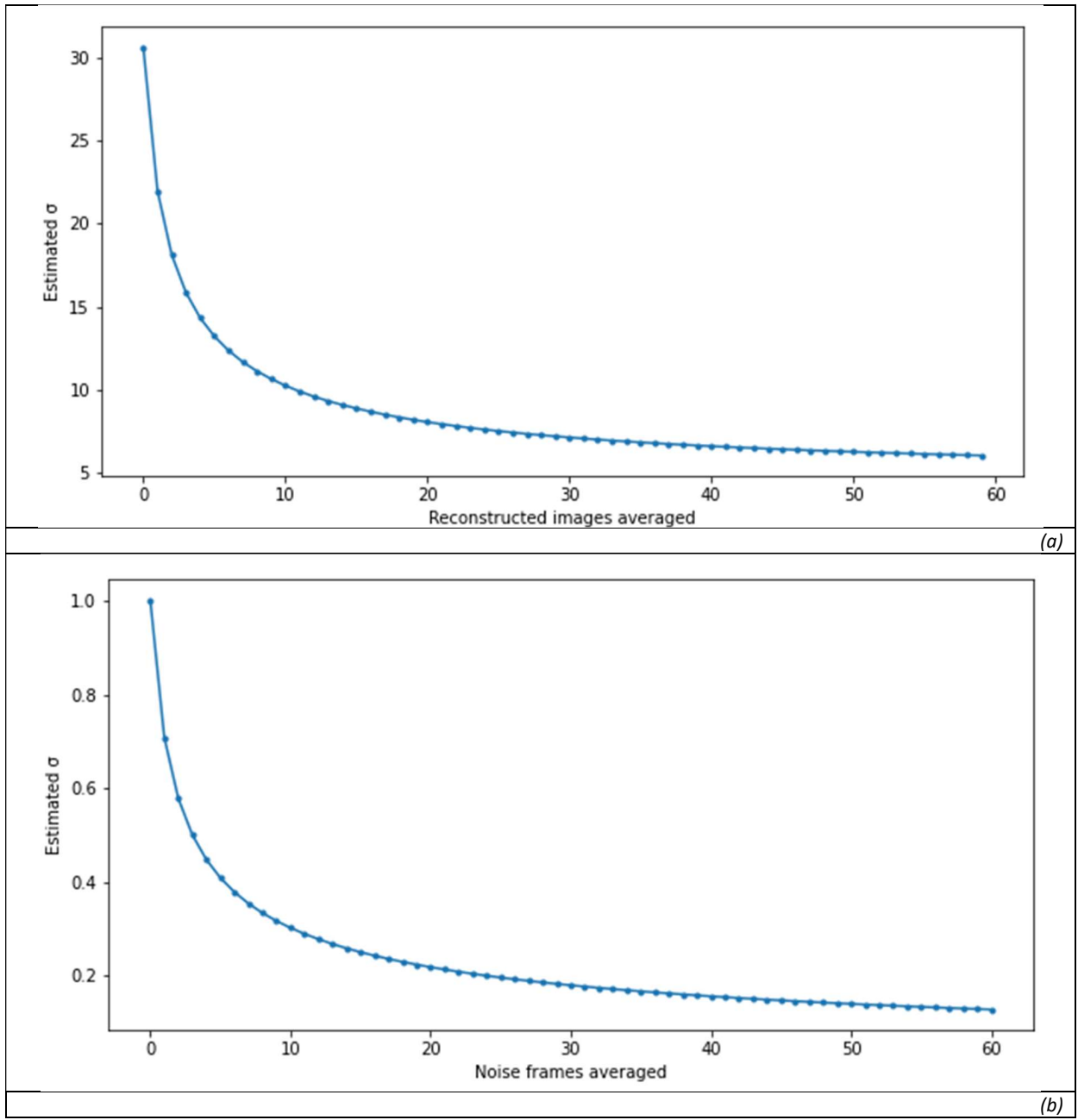

Figure 7: Estimated standard deviation ($\sigma$) from an average of multiple reconstructions. As more reconstructions are averaged together, each with different reconstructed noise, the apparent noise in the averaged image continuously decreases (a). This trend is also observed in the estimated noise from averaged frames of generated noise only (b), suggesting that this decrease in noise is simply the result of additional exposures.

Tests on experimentally measured data behave similarly to generated test data. For a very short exposure containing significant amounts of noise taken on 11-ID-B at the Advanced Photon Source for naphthalene *(fig 8a)*, the standard deviation of the noise in the original data set is calculated to be $\sigma = $ *13.5*. A single reconstruction from 30% of the pixels has noise with an estimated standard deviation of $\sigma$

= 9.66. The average of 20 of reconstructions from different subsamplings results in an estimated standard deviation of $\sigma = 3.75$ for the averaged image.

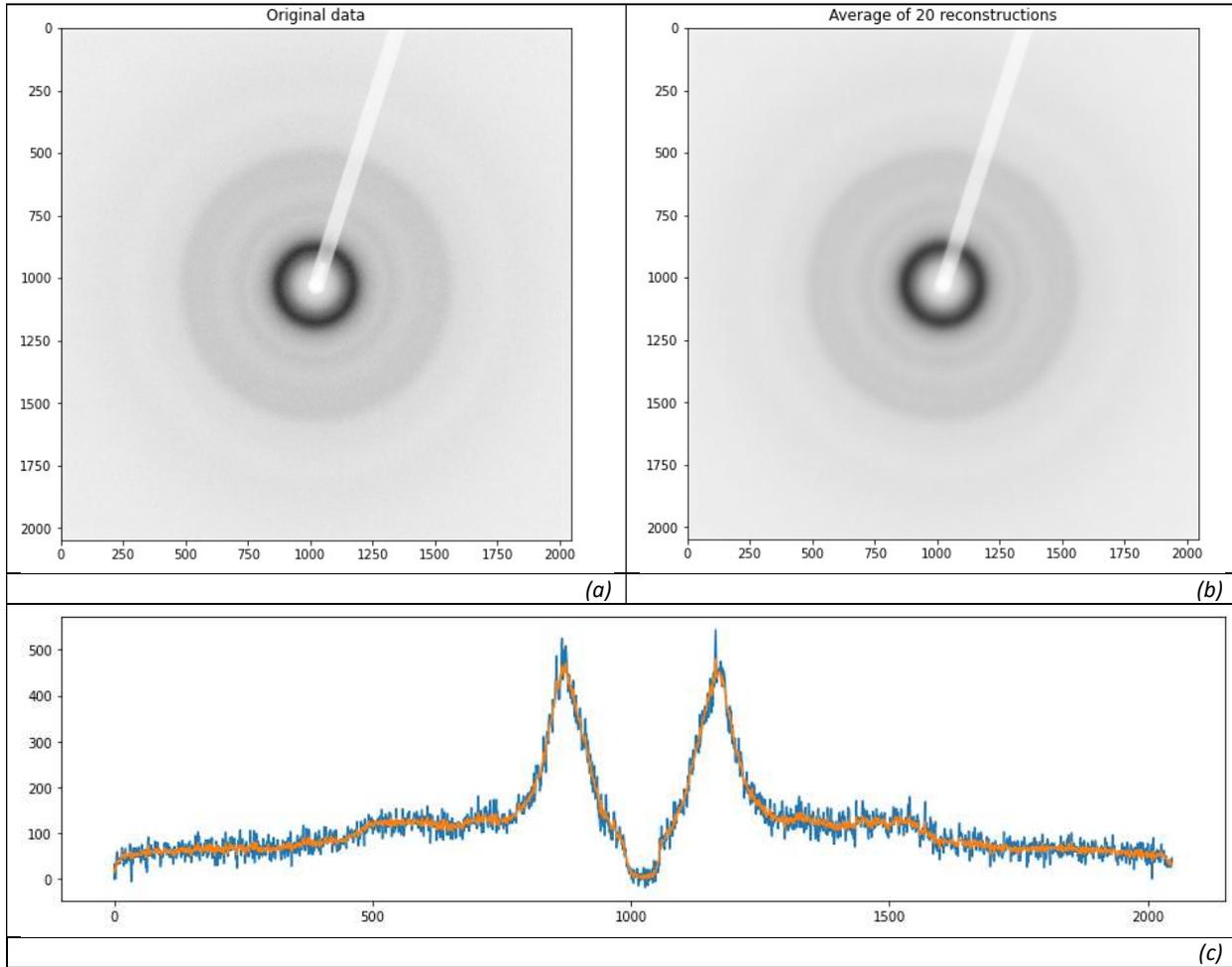

*Figure 8: An experimentally measured diffraction pattern of naphthalene (a) along an average of 20 compressed sensing reconstructions created from different randomly selected subsamplings of 30% (b). Estimated standard deviations of noise in the images are 13.5, and 3.75. A trace through row 1024 of the original and averaged image shows a significant reduction in apparent noise. Measurements were taken at 11-ID-B with a 2048x2048 Perkin Elmer detector.*

**Discussion:**

It is not currently clear if there are rules which govern the decreasing trend in the reconstruction error, though there is some practical limit to the number of useful virtual detector images that can be pulled from a single real detector measurement. For the test data after around 15 reconstructions the $\sigma$ of the noise approaches a value below $1/5^{th}$ the original noise in the real detector measurement, which is a fairly significant improvement. The computations are fairly computer intensive and at some point it may be more practical to simply take more real detector measurements. The behavior of the noise level in the average of multiple virtual detector measurements continuously decreases as more images are added (fig 7a), which is what would intuitively be expected as more repeated measurements are averaged together. Averaging multiple randomly generated images of only noise also shows an identical trend (fig 7b), suggesting that these virtual detector measurements behave similarly to independent measurements, though there appear to be some limitations which are not fully clear as of this time. This trend holds regardless of the subsampling fraction used, assuming that the subsampling fraction is sufficient to reconstruct the signal of interest. It is a additionally identical to what would be expected for a plot of the signal to noise ratio (SNR) vs the counting time of a detector, showing a linear trend when SNR is plotted vs number of exposures (SI figure 4).

Because the creation of virtual detector measurements is reliant on the incorrect reconstruction of noise, there is likely some introduced structure in the noise contained in the virtual detector measurement. It is not clear to what extent the structure

**Conclusions:**

The ability to turn one real measurement into a nearly arbitrarily large number of virtual measurements may seem counterintuitive. However, within the context of compressed sensing, only a small number of pixel positions are needed to obtain the information of interest (a detector image). That is, only a few pixels must be sampled to provide a compressed sensing measurement of the full image. A full real detector image where every pixel position is sampled thus provides a very large number of possible compressed sensing measurements. The use of compressed sensing for diffraction measurements provides an entirely new paradigm for how to collect measurements. As shown in this work, application with no changes to existing instrument detector architecture provides significant improvements in the quality of obtained measurements.

For diffraction experiments which rely on the accurate measurement of subtle variations in peak intensities and shapes, the ability to dramatically reduce noise is of significant value. Measurements for pair distribution function analysis for example, relies on the accurate capture of peak shapes and weak diffuse scattering. For diffuse scattering, which is generally many orders of magnitude weaker than bragg scattering, detector noise is often enough to prevent its measurement entirely. Samples which are easily damaged by the x-ray beam, such as (bio)molecular materials , would benefit greatly the ability to dramatically reduce noise as it would open up the ability to take shorter exposures that do not alter the sample being measured. Experiments which need high time resolution, and thus short exposures by nature, naturally also benefit from such techniques.

---